\documentclass[referee]{aa}
\usepackage{rotate}
\usepackage{graphicx}
\begin{document}

\def\ni{\noindent}
\def\be{\begin{equation}}
\def\ee{\end{equation}}
\def\lesssim{\raisebox{-0.3ex}{\mbox{$\stackrel{<}{_\sim} \,$}}}
\def\gtrsim{\raisebox{-0.3ex}{\mbox{$\stackrel{>}{_\sim} \,$}}}

\title{Drifting subpulses and inner acceleration regions in radio pulsars}

\author{Janusz Gil\inst{1,2} \and George I. Melikidze\inst{1,3} \and Ulrich
Geppert\inst{2}}

\offprints{Janusz Gil, \email{jag@astro.ia.uz.zgora.pl}}

\institute{Institute of Astronomy, University of Zielona G\'ora ,
Lubuska 2, 65-265 Zielona G\'ora, Poland \and Astrophysikalisches
Institut Potsdam, An der Sternwarte 16, 14482, Potsdam, Germany
\and Center for Plasma Astrophysics, Abastumani Astrophysical
Observatory, Al. Kazbegi ave. 2a, Tbilisi 380060, Georgia}

\abstract{The classical vacuum gap model of Ruderman \&
Sutherland, in which spark-associated subbeams of subpulse
emission circulate around the magnetic axis due to the $\vec{{\bf
E}\times{\bf B}}$ drift of spark plasma filaments, provides a
natural and plausible physical mechanism of the subpulse drift
phenomenon. Moreover, this is the only model with quantitative
predictions that can be compared with observations. Recent
progress in the analysis of drifting subpulses in pulsars has
provided a strong support to this model by revealing a number of
subbeams circulating around the magnetic axis in a manner
compatible with theoretical predictions. However, a more detailed
analysis revealed that the circulation speed in a pure vacuum gap
is too high when compared with observations. Moreover, some
pulsars demonstrate significant time variations of the drift rate,
including a change of the apparent drift direction, which is
obviously inconsistent with the ${\bf E}\times{\bf B}$ drift
scenario in a pure vacuum gap. We attempted to resolve these
discrepancies by considering a partial flow of iron ions from the
positively charged polar cap, coexisting with the production of
outflowing electron-positron plasmas. The model of such
charge-depleted acceleration region is highly sensitive to both
the critical ion temperature $T_i\sim 10^6$~K (above which ions
flow freely with the corotational charge density) and the actual
surface temperature $T_s$ of the polar cap, heated by the
bombardment of ultra-relativistic charged particles. By fitting
the observationally deduced drift-rates to the theoretical values,
we managed to estimate polar cap surface temperatures in a number
of pulsars. The estimated surface temperatures $T_s$ correspond to
a small charge depletion of the order of a few percent of the
Goldreich-Julian corotational charge density. Nevertheless, the
remaining acceleration potential drop is high enough to discharge
through a system of sparks, cycling on and off on a natural
time-scales described by the Ruderman \& Sutherland model. We also
argue that if the thermionic electron outflow from the surface of
a negatively charged polar cap is slightly below the
Goldreich-Julian density, then the resulting small charge
depletion will have similar consequences as in the case of the
ions outflow. We thus believe that the sparking discharge of a
partially shielded acceleration potential drop occurs in all
pulsars, with both positively (``pulsars'') and negatively
(``anti-pulsars'') charged polar caps.
}

\titlerunning{Drifting subpulses...}
\authorrunning{Gil, Melikidze, \& Geppert}

\maketitle

\section{Introduction}

The phenomenon of drifting subpulses has been widely regarded as a
powerful diagnostic tool for the investigations of mechanisms of
pulsar radio emission. The drifting subpulses change phase from
one pulse to another in a very organized manner, to the extent
that they form apparent driftbands of duration from several to a
few tenths of consequtive pulses. Typically, more than one drift
band appears, and the separation $P_3$ between them measured in
pulsar periods $P_1$ ranges from about 1 to about 15 (Backer
\cite{b73}; Rankin \cite{r86}; this paper). There are typically
two or three approximately equidistant subpulses in each single
pulse, separated by $P_2$ degrees of longitude. Thus, the observed
drift rate $D_0=P_2/P_3$ degrees of longitude per pulse period
$P_1$, provided that $P_3$ is alias-free or alias-corrected (e.g.
van Leeuven 2003; see also Gil \& Sendyk \cite{gs03} and
references therein for review). The subpulse intensity is
systematically modulated along drift bands, either decreasing
(typically) or increasing (seldom) towards the edge of the pulse
window. In some pulsars, however, only periodic intensity
modulations are observed, without any systematic phase change.
These pulsars were identified as those in which the line-of-sight
cuts through a beam centrally (Backer \cite{b73}), thus showing a
steep gradient of the polarization angle curve (e.g. Lyne \&
Manchester \cite{lm88}). On the other hand, the clear subpulse
driftbands are typically found in pulsars associated with the
line-of-sight grazing the beam, thus showing relatively flat
position angle curve (Backer \cite{b73}; Rankin \cite{r86}; Lyne
\& Manchester \cite{lm88}). The observed periodicities related to
patterns of drifting subpulses are independent of radio frequency,
thus excluding all frequency dependent plasma effects as a
plausible source of the drifting subpulse phenomena.

The observational characteristics of drifting subpulses described
briefly above suggest unequivocally the interpretation of this
phenomenon as a number of isolated subbeams of radio emission,
spaced more or less uniformly in the magnetic azimuth, and
rotating slowly around the magnetic axis. The most spectacular
confirmation of this interpretation was recently presented by
Deshpande \& Rankin (1999, 2001; DR99 and DR01 hereafter) and
Asgekar \& Deshpande (\cite{ad01}), who performed a sophisticated
fluctuation spectra analysis of single pulse data from PSR 0943+10
and detected clear spectral features corresponding to the
rotational behaviour of subpulse beams. If each subbeam completes
one full rotation around the magnetic axis in $\hat{P}_3$ pulsar
periods $P_1$, then $N=\hat{P}_3/P_3$ is the number of circulating
subbeams. The primary periodicity $P_3$
 is relatively easy to measure,
either by eye-inspection or by finding a corresponding frequency
$f_3=1/P_3$ in the intensity modulation spectrum (although in some
cases it requires an aliasing resolving - see Gil \& Sendyk 2003,
van Leeuwen et al. 2003, DR01, DR99). However, measuring or
estimating the circulational (tertiary) periodicity $\hat{P}_3$ is
much more difficult, since detecting a corresponding feature in
the fluctuation spectrum at low frequency $\hat{f}_3=1/\hat{P}_3$
requires an extraordinary stability of intensity patterns over a
relatively long period of time. So far, such a low frequency
feature was found only in the fluctuation spectrum of PSR 0943+10,
Asgekar \& Deshpande (2001; see their Figs.~1 and 2) detected
clear peak at $\hat{f}_3=0.027/P_1$, corresponding to
circulational period $\hat{P}_3\approx 37P_1$ (see Gil \& Sendyk
2003, for more detailed discussion). Moreover, DR99 \& DR01
detected sideband features near the high frequency feature
$f_3=1/P_3$ (separated from it by $0.027/P_1$), also clearly
associated with the rotational cycle $\hat{P}_3\sim 37P_1$ in PSR
0943+10. Van Leeuwen et al. (\cite{vl03}) analyzed drifting
subpulses in the well known pulsar PSR 0809+74 and argued that
$\hat{P}_3>150 P_1$ and $N\geq 14$ in this pulsar. Recently, Gupta
et al. (\cite{getal03}) analyzed drifting subpulses in PSR 0826-34
and found that $P_3\approx P_1$, $\hat{P}_3=14P_1$ and $N=14$ in
this pulsar (see Sect.~4.3 for some details of this analysis). We
will use the observationally deduced values of circulational
periodicities $\hat{P}_3$ of these three pulsars (and a few
others) later on in this paper in order to estimate basic
parameters of the polar cap physics.

The frequency independence of the drifting periodicities $P_3$ and
$\hat{P}_3$, as well as similarities of the drifting subpulse
patterns at different radio frequencies strongly suggest that the
radiation subbeams in the emission region reflect some kind of a
``seeding'' phenomenon at or very near the surface of the polar
cap (rather than some kind of magnetospheric plasma waves; e.g.
Kazbegi et al. 1996). As DR99 emphasize, the results of their
analysis of drifting subpulses in PSR 0943+10 appear fully
compatible with the Ruderman \& Sutherland (\cite{rs75}; RS
hereafter) ${\bf E}\times{\bf B}$ drift model, although their
analysis is completely independent of this model. In this paper we
provide further support to this natural model, whose original
version we review shortly in Appendix A.2. In Sect.~2 we discuss a
modified version of RS model, allowing a partial ion or electron
flow from the polar cap surface, coexisting with the magnetic
electron-positron pair plasma production. In Sect.~3 we discuss a
thermostatic regulation of the polar cap and estimate actual
surface temperatures. In Sect.~4 we calculate the predicted
circulational periodicities and compare them with the
observationally inferred values for a number of pulsars. Finally,
we give a summary of our results in Sect.~6.

The original version of RS model can be applied only to pulsars
with a positively charged polar cap (``pulsars'' in RS
terminology), i.e. with ${\bf\Omega}\cdot{\bf B}<0$, which is
usually considered as a deficiency of the model. We propose a
natural solution for the other ``half'' of neutron stars
(``antipulsars'' in RS terminology), with ${\bf\Omega}\cdot{\bf
B}>0$, in Sect.~2.2. RS assumed a strong binding of iron ions,
which therefore could not be released from the polar cap surface
by thermionic and/or field emission. In this paper we argue that
even if the iron ions (or electrons) are marginally bound within
the surface, then the centrifugal outflow of charges through the
light cylinder results in the creation of an acceleration region
just above the polar cap surface. The residual potential drop is
strong enough to be discharged by the magnetic creation of
electron-positron $(e^-e^+)$ pairs that form a system of isolated
plasma filaments (sparks), which in turn produce a system of
isolated plasma streams flowing along dipolar magnetic field lines
and radiating spark-associated coherent subpulse radio emission at
higher altitudes (Kijak \& Gil 1998; Melikidze et al. 2000).

The important feature of the RS model is an inevitable ${\bf
E}\times{\bf B}$ drift, which makes the spark plasma filaments
rotate about the symmetry axis of the surface magnetic field. Gil
et al. (1993) argued that this circumferential motion of sparks is
manifested by conal structure of pulsar beams (Rankin 1983, 1986).
RS adopted a pure axial symmetry of a star-centered global dipolar
field, although they implicitly assumed significantly non-dipolar
radii of curvature ${\cal R}_6\sim 1$ (see Appendix A.1.) required
by conditions of the magnetic pair production. Both the spark
characteristic dimension, as well as the distance between adjacent
sparks should be about the height $h$ of a quasi-steady vacuum gap
(RS, Gil \& Sendyk \cite{gs00}; hereafter GS00). The speed of the
${\bf E}\times{\bf B}$ drift motion around the pole is
$v_d=c\Delta E_\perp/B_s$~cm s$^{-1}$, where $\Delta E_\perp$ is
the component of the electric field caused by the charge depletion
in the acceleration region. A prominent subpulse drift can be
observed when the line-of-sight grazes the pulsar beam, which
corresponds to peripheral sparks drifting at a distance $d\sim
r_p-h$ from the pole (see Appendix A.2. for details). Each spark
completes one full rotation around the magnetic axis in
$\hat{P}_3=2\pi d/v_d$~seconds (called the tertiary periodicity).
According to Eqs.~(A.3) and (A.4), for $d\approx r_p-h$ and
$\eta=1$ (pure RS vacuum gap) the tertiary periodicity
$\hat{P}_3/P_1=[(r_p/h)-1]$ and the azimuthal drift rate
$D_r=360^\circ/\hat{P}_3=360^\circ/[(r_p/h)-1]$. It is important
to emphasize that the value of $D_r$ can be deduced from
observations of drifting subpulses only if the value of
$\hat{P}_3$ can be measured/estimated. On the other hand, $D_r$
can be theoretically estimated if the value of the ratio $r_p/h$
is known. In the case of PSR 0943+10, the value of $r_p/h\sim 7$
(see GS00 and Gil et al. 2002b; GMM02b hereafter;) and
$\hat{P}_3=37.35~P_1$. Thus, the vacuum drift (RS) periodicity
$\hat{P}_3\sim 6P_1$ is about six times shorter than the observed
value $\hat{P}_3\approx 37P_1$, and, consequently, the drift rate
$D_r=360^\circ/\hat{P}_3$ is about 6 times too high (see also Gil
\& Sendyk \cite{gs03}). In PSR 0809+74, which is another pulsar
for which $\hat{P}_3$ could be estimated, this discrepancy is much
larger. In fact, van Leeuwen et al. 2002 demonstrated that the
observationally deduced value of $\hat{P}_3$ exceeds $150P_1$,
while the RS value of $\hat{P}_3$ is about $5P_1$. Similarly, in
PSR 0826-34 the RS model gives $\hat{P}_3=5P_1$, while Gupta et
al. (2003) deduced from the analysis of drifting subpulses that
$\hat{P}_3\simeq 14P_1$ in this pulsar.

Therefore, in pulsars for which the azimuthal (intrinsic) drift
rates $D_r$ can be measured/estimated (see Table 1), they turn out
to be a few to several times lower than those predicted from RS
model. In other words, the pure vacuum gap drift is too fast as
compared with observations. Moreover, in some pulsars a time
variable drift rate is observed, including reversals of apparent
drift direction in a few cases. These observational features are
inconsistent with the RS model, which otherwise provides a quite
natural and plausible physical mechanism of the subpulse drift
phenomenon (not to mention that this is the only quantitative
model that can be compared with observations). In this paper we
attempt to resolve these discrepancies within a more general model
of the inner acceleration region, involving a partial flow of iron
ions (or electrons) due to the thermal emission from the polar cap
surface, heated to high temperatures by sparking discharges. Such
generalization of the pure vacuum gap model of RS was first
proposed by Cheng \& Ruderman (1980, CR80 hereafter; see also Usov
\& Melrose 1995, 1996). However, CR80 suggested that even with
ions included in the flow, the conditions above the polar cap are
close to a pure vacuum gap. We, on the contrary, argue in this
paper on both theoretical and observational grounds that the
quasi-stationary discharge conditions can be established, even if
the ion (electron) flow exceeds 95\% of the Goldreich \& Julian
(1969; GJ hereafter) charge density. Nevertheless, the remaining
acceleration potential drop is high enough to discharge through a
system of sparks, as originally proposed by RS. The important
difference is that the ion (electron) flow may strongly reduce the
${\bf E}\times{\bf B}$ drift-rate, to a level comparable with
observationally deduced values. The time dependent shielding can
result in a time variability of the observed drift-rate, including
the apparent reversals of the drift direction. The latter effect
can occur if the natural sampling rate (once per pulsar period) is
too slow with respect to the drifting subpulses variability and
results in an aliasing phenomenon. In fact, the apparent reversals
of subpulse drift direction can be explained by small variations
(a few percent of the mean value) of the drift rate, which cause
the $P_3$ value to fluctuate around the relevant Nyquist boundary.

\section{Critical temperatures}

The value of the charge density above the polar cap heated by
discharge bombardment is limited by the co-rotational GJ value.
Since the number density of iron ions or electrons in the neutron
star crust is many orders of magnitude larger than corotational
values above the surface, then a thermionic emission from the
polar cap surface is not simply described by the usual condition
$\varepsilon_c\approx kT_s$, where $\varepsilon_c$ is the cohesive
energy and/or work function, $T_s$ is the actual surface
temperature and $k$ is the Boltzman constant. Below we consider
pulsars with positively (ions) and negatively (electrons) charged
polar caps separately.

\subsection{Iron critical temperature $T_i$}

In neutron stars with positively charged (${\bf\Omega}\cdot{\bf
B}<0$) polar caps the outflow of iron ions is limited by
thermionic emission and determined by the surface-binding
(cohesive) energy $\varepsilon_c$. Let us consider, following the
results of CR80, a general case of a pulsar inner accelerator in
the form of a charge depletion region rather than a pure vacuum
gap. According to their Eq.~(8) the outflow of iron ions can be
described in the form \be \frac{\rho_i}{\rho_{GJ}}\approx
exp\left(30-\frac{\varepsilon_c}{kT_s}\right) ,\label{rojon}\ee
where $\rho_i\leq \rho_{GJ}$ is the charge density of outflowing
ions. At the critical temperature \be
T_i=\frac{\varepsilon_c}{30k} \label{tejon} ,\ee the ion outflow
reaches the maximum value $\rho=\rho_{GJ}$ (Eq.~(A.1)) permitted
by the force-free magnetospheric condition. The numerical
coefficient equal to 30 in Eqs.~(\ref{rojon}) and (\ref{tejon}) is
determined from the tail of the exponential function with an
accuracy of about 10\%. Thus, for a given value of the cohesive
energy $\varepsilon_c$, the critical temperature $T_i$ is also
estimated within an accuracy of about 10\%. Different values of
the cohesive energy $\varepsilon_c$ obtained by different authors
lead to different values of critical temperatures. According to
calculations of Abrahams \& Shapiro (\cite{as91}) \be T_i=(6\times
10^5)b^{0.73}(P_1\dot{P}_{-15})^{0.36} ~{\rm K}, \label{teia}\ee
while calculations of Jones (\cite{j86}) give values five times
lower \be T_i=(1.2\times
10^5)b^{0.7}(P_1\dot{P}_{-15})^{0.36}~{\rm K}, \label{teijot}\ee
where the parameter $b=B_s/B_d$ is described in Appendix~A.1. (see
also Eqs.~(\ref{rojon}) and (\ref{tejon}) in Gil \& Melikidze
2002; GM02 hereafter).

Below the critical temperature $T_i$ the charge-depleted
acceleration region will form, with the accelerating potential
drop $\Delta V=\eta\Delta V_{max}$, where $\Delta V_{max}$ is the
pure vacuum gap potential drop (Eq.~(A.2)),
 and the shielding factor can be
defined/expressed in the form \be
\eta=1-\rho_i/\rho_{GJ}=1-exp[30(1-T_i/T_s)] .\label{eta}\ee

\subsection{Electron critical temperature $T_e$}

Let us now consider pulsars with negatively charged polar caps
(${\bf\Omega}\cdot{\bf B}>0$), called ``antipulsars'' by RS. It is
conventionally assumed that in such case a stationary free flow of
electrons with the corotational GJ density (Eq.~(A.1)) exists.
This is the so-called space charge limited flow (SCLF), in which
the accelerating potential drop arises due to the dipolar field
line curvature and/or inertia (Arons \& Sharleman \cite{as79};
Arons \cite{a81}). Such a free flow requires that the electron
work function $w$ is completely negligible. However, it is known
that $w$ is of the order of 1 keV (see Eq.~(C.6)), which is
comparable with the ion cohesive energy $\varepsilon_c$ (Eq.~(2)).
It is therefore possible that the electron flow is determined
mainly by thermoemission. If, similarly to the ions case, the
electron flow is slightly below the GJ value, then the effective
potential drop just above the polar cap will be dominated by a
small depletion of negative charge. This can happen if the actual
surface temperature $T_s$ is slightly smaller than the critical
electron temperature $T_e$ (Eq.~(\ref{tee})). In fact, as
demonstrated by Usov \& Melrose (1996; see also Appendix C in this
paper), the flow of electrons ejected from the polar cap surface
by thermionic emission provides the GJ charge density if
$T_s>T_e\simeq 0.04w/k$. One can therefore write electron
analogues of Eqs.~(\ref{tejon}) and (\ref{eta}) in the form \be
T_e=\frac{w}{25k} ,\ee and \be
\eta=1-\rho_e/\rho_{GJ}=1-exp[25(1-T_e/T_s)] ,\label{eta2}\ee
where $\rho_e$ is the charge density of thermionic electrons,
$T_e$ is the critical surface temperature and $T_s\lesssim T_e$ is
the actual surface temperature.

The critical electron temperature $T_e$ is determined in terms of
basic pulsar parameters by Eq.~(\ref{cetee}) in Appendix C. Using
Eqs.~(\ref{bees}) and (\ref{bede}) we can rewrite this equation in
the form \be T_e\simeq (5.94\times
10^5)b^{0.4}P_1^{0.16}\dot{P}^{0.2}_{-15}~{\rm K}. \label{tee}\ee

As one can see, the values of the critical electron temperature
$T_e$ are close to the critical ion temperature $T_i$ obtained by
Abrahams \& Shapiro (\cite{as91}). The observational estimates of
polar cap surface temperatures $T_s$ based on first results from
XMM satelite indicate values above $10^6$~K (Becker \& Aschenbach
\cite{ba02}). Therefore, the parameter $b$ in Eqs.~(\ref{teia}),
(\ref{teijot}) and (\ref{tee}) must be considerably larger then
unity, implying a strong non-dipolar surface magnetic field (Eqs.
(A.2) and (A.3)).

The original RS vacuum gap model is known to have a fundamental
problem with the cohesive (binding) energy $\varepsilon_c$, which
is apparently too low to bind the $^{56}_{26}$Fe ions in the
uppermost layer of the polar cap surface (for review see Usov \&
Melrose 1995). However, Gil \& Mitra (2001; GM01 hereafter) and
GM02 argued recently that RS-type vacuum gap can form above a
positively charged polar cap, provided that the surface magnetic
field $B_s=bB_d$ (Eqs.~(A.2) and (A.3)) is very strong (about
$10^{13}$~ G) and non-dipolar in nature (radius of curvature
${\cal R}\leq 10^6$~ cm). In the partially shielded acceleration
region $(\eta<1)$, the actual surface magnetic field can be a few
times lower, although still much higher than the conventional
dipolar field $B_d$ (Table 1). In fact, one can show that the
minimum surface magnetic field necessary for the formation of a
pure vacuum gap (with $\eta=1$) obtained by GM02 (their Eqs. (8)
and (14) and Fig. 1) should now be multiplied by a factor
$\eta^{0.57}$. For six pulsars listed in Table 1, the average
value of this factor is about 0.3 and the average value of the
parameter $b=B_s/B_d$ is about 4 and the average value of
$B_d=3.3\times 10^{12}$~G, while the average value of
$B_s=8.7\times 10^{12}$~G.

\section{Thermostatic regulation of the actual surface temperature $T_s$}

Let us consider a quasi-equilibrium state when cooling by
radiation balances heating due to electron bombardment of the
polar cap surface
 \be \sigma T_s^4=\eta\gamma_{acc}m_ec^3n_{GJ} ,\label{sigmate}\ee
 where the shielding factor $\eta$ is determined by Eqs.~(\ref{eta})
 or (7), the Lorentz factor
 $\gamma_{acc}$ is determined by Eq.~(A.12)\footnote{Although the Lorentz
factor
 $\gamma_{acc}$ was calculated within a framework of the NTVG-ICS model of GM01 (see
 also GM02 and Appendix A), its value is not strongly dependent on the adopted model of
 acceleration region (therefore $T_s$ is not strongly dependent on the
 particular inner gap acceleration model).},
$n_{GJ}=\rho_{GJ}/e=1.4\times 10^{11}b(\dot{P}_{-15}/P)^{0.5}~{\rm
cm}^{-3}$, $m_e$ is the electron mass, $e$ is the elementary
charge, $c$ is the speed of light and $\sigma$ is the
Stefan-Boltzman constant. It is straightforward to obtain the
expression for the quasi-equilibrium surface temperature in the
form \be T_s=4.34\times 10^6P_1^{-1/7}{\cal R}_6^{2/7}\eta^{3/7}~
{\rm K}.\label{te6}\ee Inverting this equation we obtain the
expression for the shielding factor $\eta$ in terms of the surface
temperature $T_s$ \be \eta=3.25\times
10^{-2}T_6^{7/3}P_1^{1/3}{\cal R}_6^{-2/3} ,\label{eta1}\ee where
$T_6=T_s/10^6$~K. This expression describes the shielding factor
$\eta$, in terms of the balance Eq.~(\ref{sigmate}) independently
of the general definition (Eqs.~(\ref{eta}) or (\ref{eta2})).
Since Eq.~(\ref{eta1}) and Eq.~(\ref{eta}) or Eq.~(\ref{eta2})
have to be satisfied simultaneously, then the actual surface
temperature $T_s$ will be thermostatically self-regulated within a
narrow range around the quasi-equilibrium value (Eq.~(\ref{te6})).
In fact, a slight decrease of $T_s$ in Eq.~(\ref{eta}) causes an
increase of the shielding factor $\eta$ (due to a smaller number
of thermionic ions). This in turn causes an increase of $T_s$ in
Eq.~(16), due to a less shielded accelerating potential drop and
more intense heating by discharge bombardment.

\begin{figure}
\includegraphics[scale=0.9, angle=0]{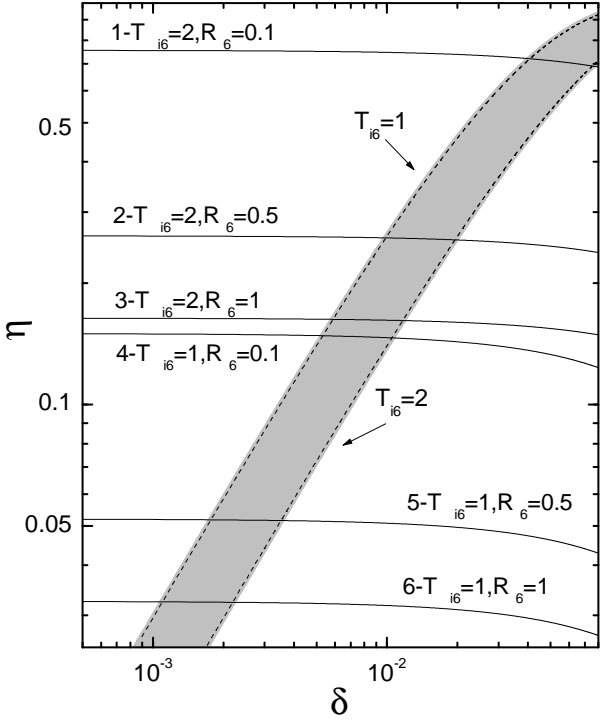}
\caption{ Potential shielding factor $\eta$ calculated from
Eq.~(\ref{eta}) - two dashed lines, and from Eq.~(\ref{eta1}) -
six solid lines numbered from 1 to 6, versus the temperature
difference $\delta=(T_i-T_s)/10^6$~K, where $T_i$ is the ion
critical temperature (Eq.~(\ref{tejon})) and $T_s<T_i$ is the
actual surface temperature.}
\end{figure}

Figure 1 shows values of the shielding factor $\eta$ calculated
from Eqs.~(\ref{eta}) represented by two dashed lines, and
(\ref{eta1}), represented by six numbered solid lines. We
introduced a parameter $\delta =T_{i6}-T_6$ shown on the
horizontal axis, which describes a difference between the iron
critical temperature $T_i=T_{i6}\times 10^6$~K and the actual
surface temperature $T_s=T_6\times 10^6$~K. Using this parameter
we can rewrite Eq.~(\ref{eta}) in the form $\eta =1-exp\left
\lbrack -30\delta/(T_{i6}-\delta)\right\rbrack$, and two dashed
lines presented in Fig.~1 correspond to limiting values $T_{i6}=$1
and 2, respectively\footnote{Our conclusions are not very
sensitive to these limiting values, provided that they are not
much different from 1 and 2 (see Sect.~4).}. Similarly,
Eq.~(\ref{eta1}) can be rewritten in the form $\eta =3.25\times
10^{-2}(T_{i6}-\delta)^{7/3}{\cal R}_6^{-2/3}P_1^{-1/3}$, and six
numbered lines presented in Fig.~1 correspond to different
(indicated) combinations of $T_{i6}$ and ${\cal R}_6$ values
($P_1=1$ was used in all cases). The actual values of $\eta$
correspond to the shadowed region limited by the two dashed lines
and the two numbered lines 1 and 6 from the top and the bottom,
respectively. If the iron critical temperature $T_i$ is not much
lower than $10^6$~K and (\ref{teia})) and ${\cal R}_6$ is not much
smaller than 0.1, then the value of $\eta$ should lie in the range
0.03-0.8. This suggests that the actual value of the parameter
$\delta$ should range between a fraction of a percent to several
percent at most. We therefore conclude that a difference between
$T_i$ and $T_s$ (or $T_e$ and $T_s$) should be of the order of
$10^4$~K. In the next section we constrain both $T_i$ (or $T_e$)
and $T_s$ from the observationally deduced drift-rates for a
number of pulsars. The inferred surface temperature $T_s$ and the
difference between $T_i$ (or $T_e$) and $T_s$ agree very well with
a general estimate obtained here.

The model of a charge-depleted acceleration region described by
Eqs.~(\ref{rojon}) and Eq.~(\ref{eta}) for ions and
Eq.~(\ref{eta2}) for electrons is highly sensitive to both
$T_{i,e}$ and $T_s$ temperatures, where $T_{i,e}$ means either
$T_i$ (Eqs.~(\ref{teia}), (\ref{teijot})) or $T_e$
(Eq.~(\ref{tee})). The inequality $T_s<T_{i,e}$ is often used as a
thermal condition for the ``vacuum gap formation'' (GM00; GM02;
Abrahams \& Shapiro \cite{as91}; Usov \& Melrose \cite{um95},
\cite{um96}) and its exact meaning is worth understanding. When
the surface temperature $T_s$ is only 10\% below $T_{i,e}$, then
$\eta\approx 0.95$, that is $\rho_{i,e}$ is only a few percent of
the corotational GJ charge density and the situation is close to
the pure vacuum case of RS (see discussion below Eq.~(8) in CR80
for details). If this is the case, then the potential drop $\Delta
V\approx\Delta V_{max}$ is being screened mostly due to the
intense production of $e^-e^+$ pairs, that is, the total charge
density $\rho_{t}\leq\rho_{GJ}$ (equality holds when a spark
plasma filament is fully developed), where $\rho_t=\rho_i+\rho_+$
or $\rho_t=\rho_e+\rho_-$, and $\rho_+$ and $\rho_-$ are charge
densities of positrons and electrons produced in $e^-e^+$ pairs in
spark discharges, respectively.
 The back flow of high density electrons or positrons
 accelerated to relativistic energies by the vacuum potential
drop, heats strongly the surface beneath the sparks. As a result,
the surface temperature $T_s$ increases, which may intensify the
thermionic ion/electron outflow, additionally screening the
potential drop to the level $\Delta V=\eta\Delta V_{max}$. At this
stage, the cooling process should slightly dominate, because of
weaker heating due to the back flow of lower density
$\rho_\pm=\eta\rho_{GJ}$ and lower energy $\gamma=\eta\Delta
V_{max}/m_ec^2$ electrons/positrons. Thus, the actual surface
temperature should be thermostatically self-regulated. If heating
and cooling are nearly in balance, then the quasi-equilibrium
surface temperature $T_s$ is very close to $T_i$ (Fig.~1), with
the difference being less than about 1\%. Thus, contrary to the
suggestion of CR80 (see discussion below their Eq.~(8)), the
conditions close to a pure vacuum gap will never be established in
the quasi-equilibrium stage. Nevertheless, the potential drop
arising due to a small charge depletion above the polar cap is
high enough for the non-stationary sparking scenario to be
realized in the way similar to that envisioned by RS.

The sparking discharge terminates if the total charge density
$\rho_t=\rho_{i,e}+\rho_\pm$ reaches the GJ value (Eq.~(1)), where
$\rho_{i,e}$ is described by Eqs.~(5) and (\ref{eta2}),
respectively, and time dependent $\rho_{\pm}(t)$ is the charge
density of produced electrons/positrons. A slight adjustment to
$\rho_{i,e}$ is also expected in the final stage of a spark
development due to the surface temperature variations. More
exactly, the discharge terminates when the screened potential drop
$\Delta V=(1-\rho_t/\rho_{GJ})\Delta
V_{max}=(\eta-\rho_\pm/\rho_{GJ})\Delta V_{max}$ falls below the
threshold value $\Delta V_{min}=\gamma_{min}(mc^2/e)\approx
2.5\times 10^3T_6^{1/3}{\cal R}^{1/3}(mc^2/e)\approx 5\times
10^8$~V, inhibiting further $e^-e^+$ pair production (see
Appendix~A.1. for determination of $\Delta V_{max}$ and
$\gamma_{min}$). Since $\Delta V_{max}=\Delta V_{ICS}\sim
(10^{11}-10^{12})$~V (see Eq.~(11) in GM02 or Eq.~(A.6) in this
paper), then $\Delta V_{min}/\Delta V_{max}\sim
10^{-3}\approx(\eta-\rho_\pm/\rho_{GJ})$. Therefore, the discharge
terminates if $\rho_\pm=\eta\rho_{GJ}$, which can be a small
fraction of GJ density. In fact, the condition
$\gamma_{acc}>\gamma_{min}$ leads to a lower limit $\eta>
10^{-4}b{\cal R}_6^{-1}T_6$, where $\gamma_{acc}$ is determined in
Eq.~(A.15). Since the product $b{\cal R}_6^{-1}$, describing small
scale anomalies of the surface magnetic field should be about 10,
the absolute lower limit for $\eta$ is about 0.001. It should be
obvious from the above considerations that the sparking discharge
can proceed in a similar way to that envisioned by RS, even if the
parameter $\eta$ is very low, say below 0.01. In reality, however,
one should expect that typically $\eta$ is of the order of 0.1
(Fig.~1) and thus $\rho_{i,e}\sim 0.9\rho_{GJ}$. Thus, one can say
that the sparking discharge begins at the partially shielded
potential drop $\Delta V\sim 0.1\ \Delta V_{max}\sim
10^{10}-10^{11}$~V and terminates when this potential drop is
largely reduced by about two orders of magnitude (due to the
screening by cascading production of $e^-e^+$ pairs).

At the final stage of a spark development, when heating is most
intense, the maximum polar cap temperature that can be achieved is
$T_s=T_{i,e}$ and the spark discharge is terminated. Then, in the
absence of heating from a pair production mechanism, the surface
temperature should drop fast enough to enable the next discharge
to recur within a sufficiently short amount of time. The e-folding
cooling time $\tau_{cool}$ is estimated in Appendix B.2. as being
of the order of 1 $\mu$s (Eq.~B.14). Thus, the surface temperature
can drop by a factor of $e$ in a time interval of the order of
$\sim \mu$s. However, in our scenario it is only required that the
temperature drops by a few to several percent, and thus a
corresponding cooling time is at least 10 times shorter. This is
comparable with a gap emptying time (or transit time) $t\sim h/c$,
which is of the order of 100 ns or less (e.g. Asseo \& Melikidze
\cite{am98}, GS00). On the other hand, the heating time scale due
to electron/positron bombardment is about $(10-40)h/c$ (RS, see
also Appendix B.3.). Therefore, sparking discharges should be
easily able to cycle on and off on a natural thermal timescales,
comparable to those described by RS model of a spark development
(exponentiation of charge density from $\rho\approx 0$ to
$\rho=\rho_{GJ}$).

\section{Subpulse drift in pulsars}

We now compare the observationally deduced drift-rates with
theoretical predictions of the ${\bf E}\times{\bf B}$ drift model
determined by Eqs.~(A.13--A.15). The input parameters and the
observationally deduced parameters are summarized in Table 1. The
most important are: the drift periodicity $P_3$ and the
circulational (tertiary) periodicity $\hat{P}_3$, which are
related to each other by the number of circulating sparks
$N=\hat{P}_3/P_3$. If necessary/possible these values are
corrected for aliasing effects. In the first three cases listed in
Table 1 we rely on a robust analysis of real observational data,
while in the next three cases (PSRs 2303+30, 2319+60 and 0031-07)
we deduce drift parameters by reproducing the actual pulse
sequences using the simulation model and fitting the free model
parameters to the observed values (see GS00 for details).

\subsection{PSR 0943+10}

This pulsar was extensively analyzed and interpreted recently by
DR99, DR01, Asgekar \& Deshpande (2001) and Gil \& Sendyk (2003).
These authors revealed clearly that drifting subpulses in this
pulsar result from a system of $N=20$ subbeams (sparks)
circulating around the magnetic axis at the periphery of the
pulsar beam (polar cap). Each subbeam (spark) completes one full
circulation in $\hat{P}_3=37.35P_1$, therefore
$P_3=\hat{P}_3/N=1.87P_1$, where $P_1$ is the pulsar period. These
precise estimates\footnote{The precise values of
$\hat{P}_3=(37.35\pm 0.52)P_1$ and $P_3=(1.87\pm 0.026)P_1$ were
given by DR99 and DR01 (at 430 and 111 MHz). Asgekar \& Deshpande
(2001) analyzing the 35-MHz observations of this pulsar confirmed
that $\hat{P}_3\sim 37P_1$. However, Rankin et al. (2003)
analyzing 103/40-MHz Pushchino observations demonstrated that both
$\hat{P}_3$ and $P_3$ values can slightly vary between different
observing session, with differences being of the order of a few
percent.} were possible due to the successful resolving of
aliasing near the Nyquist frequency $f_3\sim 0.5/P_1$ (or $P_3\sim
2P_1$).

At the periphery of the polar cap the circulation distance
$d\approx r_p-h$, where $r_p$ is the polar cap radius
(Eq.~(\ref{erpe})) and $h$ is the height of the acceleration
region (e.g. Eqs.~(\ref{haics}) and (\ref{haics1})). Thus equation
(\ref{pe3}) gives in this case
$\hat{P}_3=\eta^{-1}P_1[(r_p/h)-1]$. In order to compare this
equation with the observed value $\hat{P}_3=37.35P_1$, we have to
estimate the complexity parameter $r_p/h$, representing the ratio
of the polar cap radius to the characteristic spark dimension
${\cal D}\sim h$ at the polar cap periphery (GS00). First, let as
assume after RS that the typical distance between adjacent sparks
at the periphery of the polar cap is also about $h$. If this is
the case, then $2\pi d=2\pi(r_p-h)=2Nh$, which leads to
$r_p/h-1=N/\pi=20/\pi$ or $r_p/h\approx 7.4$ (see also GMM02b who
obtained $r_p/h=7.34\pm 0.8$ for this pulsar; in consistence with
approximate Eq.~(11) in GS00, which gives $r_p/h=6.8$). Therefore,
with $r_p/h=7.4$ we obtain $\hat{P}_3=6.4P_1/\eta=37.35P_1$, which
gives the value of the shielding parameter $\eta=0.17$, that can
be compared with theoretical predictions using Eq.~(\ref{eta}) or
(\ref{eta2}) for partial electron flow of ions or electrons,
respectively.

The actual quasi-equilibrium surface temperature is described by
Eq.~(\ref{eta1}), which for $\eta=0.17$ and ${\cal R}_6=1$ gives
$T_s=1.646\times 10^6$~K. On the other hand, from Eqs.~(\ref{eta})
and (\ref{eta2}) we obtain $T_i/T_s=1.00625$ and $T_e/T_s=1.0075$,
which gives critical iron $T_i=1.656\times 10^6$~K and electron
$T_e=1.658$~K temperatures, respectively. Note that $\Delta
T=T_i-T_s\sim T_e-T_s\sim 10^4$~K, in agreement with the general
prediction made in Sect.~3. The comparison of $T_i$ with
Eq.~(\ref{teia}) and (\ref{teijot}) for iron critical temperatures
and $T_e$ with Eq.~(\ref{tee}) for electron critical temperature
implies that $b=2.8$, 29.7 and 18 or $B_s\simeq 10^{13}$~G,
$B_s=11.2\times 10^{13}$~G and $B_s=7\times 10^{13}$~G,
respectively. As one can see, the calculations of the cohesive
energy by Jones (\cite{j86}) represented by Eq.~(\ref{teijot})
lead to values of $B_s$ largely exceeding the critical quantum
field $B_q=4.4\times 10^{13}$~G, which is believed to represent
the photon splitting limit above which the plasma necessary for
generation of pulsar radio emission can not be produced (e.g.
Zhang \& Harding 2002 and references therein). On the other hand,
calculations of Abrahams \& Shapiro (\cite{as91}) for the iron
critical temperature represented by Eq.~(\ref{teia}) give a quite
suitable value of the surface magnetic field $B_s\sim 10^{13}$~G
(Table 1). The electron case represented by Eq.~(\ref{tee}) can be
ruled out, since the corresponding $B_s=7\times 10^{13}$~G exceeds
largely the photon splitting limit.

\subsection{PSR B0809+74}

This pulsar was recently analyzed by van Leeuwen et al.
(\cite{vl03}), who argued that the number of circulating beams
$N\geq 15$ and the circulation period $\hat{P}_3>150P_1$. Assuming
again that the observed drifting subpulses in this pulsar
correspond to the periphery of pulsar beam (polar cap) we have
$\hat{P}_3=\eta^{-1}P_1[(r/h)-1]$ and $r_p/h=(N/\pi)+1\gtrsim 5.8$
(see Sect.~4.1). Thus, the shielding factor $\eta>0.032$ and using
Eq.~(\ref{eta}) we obtain $T_i/T_s=1.001$, while Eq.~(\ref{eta2})
gives $T_e/T_s=1.0013$. The actual surface temperature can be
estimated from Eq.~(\ref{eta1}), which for $\eta=0.032$ and ${\cal
R}_6=1$ gives $T_s=0.955\times 10^6$~K. Therefore we have
$T_i=0.956\times 10^6$~K and $T_e=0.957\times 10^6$~K,
respectively (notice that in this case $\Delta
T=T_i-T_s=T_e-T_s\sim 2\times 10^3$~K). These values can now be
compared with Eqs.~(\ref{teia}), (\ref{teijot}) and (\ref{tee}),
which give $B_s=4\times 10^{12}$~G (less than the minimum value
$0.1B_q$ required by near-threshold conditions described in
Appendix A.1.), $B_s=3\times 10^{13}$~G (too close to $B_q$) and
$B_s=7\times 10^{12}$~G, respectively (OK). The latter value of
the surface magnetic field is very suitable, suggesting that PSR
0809+74 is a good candidate for a drifting subpulse pulsar with a
positively charged polar cap (antipulsar in the terminology of
RS).

\subsection{PSR B0826-34}

The drifting subpulses in this remarkable pulsar were first
observed by Biggs et al. (\cite{betal85}). The average profile is
very broad and consist of the main-pulse (MP) and the interpulse
(IP). The single pulse emission occurs practically at every
longitude of the 360 degrees pulse window, implying that this
pulsar is an almost aligned rotator (i.e. the inclination angle
$\alpha$ between the rotation and magnetic axes is very small). At
645 MHz Biggs et al. (\cite{betal85}) revealed 5 to 6 bands of
drifting subpulses, swinging across the MP window. Individual
subpulses show a variable drift rate, including changes of the
apparent drift direction approximately every 100 pulses (see
Fig.~2 in Biggs et al. 1985). Such patterns of drifting subpulses
are inconsistent with the ${\bf E}\times{\bf B}$ drift model,
unless the observed effects result from aliasing and small changes
in the drift-rate causing $P_3$ to fluctuate around the relevant
Nyquist boundary (as in the case of PSR 2303+30; see Fig.~3 in
GS00). Recently Gupta et al. (\cite{getal03}) attempted to resolve
the possible aliasing in PSR 0826-34. They reobserved this pulsar
at 318 MHz using the GMRT and obtained a sequence of 500 good
quality single pulses. At this frequency 6 to 7 bands of drifting
subpulses appeared, behaving similarly to those 5-6 bands observed
by Biggs et al. (\cite{betal85}) at a higher frequency. The
careful analysis revealed that subpulse-associated sparks
circulate always in one direction consistent with the ${\bf
E}\times{\bf B}$ drift model, but the observed patterns are
determined by the aliasing corresponding to $P_3\sim P_1$ (thus
around the fluctuation frequency $f_3\sim 1/P_1$).

The observed longitudinal drift rate $D_0=d\varphi/dt=P_2/P_3$
corrected for aliasing $(P_3\sim P_1)$ is $D_0\approx P_2/P_1$,
where $P_2\approx 26^\circ$ is the typical distance between
adjacent subpulses, where $\varphi$ is the observed subpulse
longitude. The circulational periodicity
$\hat{P}_3=360^\circ/D_r$, where $D_r=d\chi/dt$ is the intrinsic
azimuthal drift rate (see Appendix A.2.). In the case of an almost
aligned rotator $(\alpha\sim 0^\circ)$ one can write $D_r\sim
D_0=26^\circ/P_1$ and thus
$\hat{P}_3\approx(360^\circ/26^\circ)P_1=13.85P_1$. Therefore, in
PSR 0826-34 there are $N=\hat{P}_3/P_3\approx\hat{P}_3/P_1=14$
sparks circulating around the pole at the rate $D_r\approx
26^\circ/P_1$.

The detailed analysis of single pulses performed by Gupta et al.
(\cite{getal03}) revealed also that sparks contributing to the MP
emission of B0826-34 circulate at $d\sim 0.5r_p$ (the middle of
the polar cap), while the IP emission occupies the periphery of
the polar cap. Thus, according to Eq.~(\ref{pe3}) with $s=0.5$,
the circulation period $\hat{P}_3=\eta^{-1}P_1(0.5r_p/h)^2$, which
can be compared with the observationally deduced value
$\hat{P}_3\sim 14P_1$. To estimate the value of $(r_p/h)$ let us
notice that in this case $2\pi d=2N{\cal D}$, where $d\sim 0.5r_p$
and ${\cal D}\sim 0.5h$ is the characteristic spark dimension in
this region of the polar cap\footnote{We assume here that ${\cal
D}\sim h$ at the polar cap boundary $(d\sim r_p)$ but due to
convergence of the planes of field lines (driving the spark
avalanche) towards the pole ${\cal D}\sim 0.5h$ in the middle of
the cap $(d\sim 0.5p)$.}. Therefore $r_p/h=N/\pi=14/\pi\sim 4.5$,
thus $\hat{P}_3=5\eta^{-1}P_1\approx 14P_1$ and, in consequence,
the shielding factor $\eta=0.357$. Now using Eqs.~(\ref{eta}) and
(\ref{eta2}) we obtain $T_i/T_s=1.0147$ and $T_e/T_s=1.0188$,
respectively. The actual surface temperature can be estimated from
Eq.~(\ref{te6}), which for ${\cal R}_6=1$ and $\eta=0.36$ gives
$T_s=2.565\times 10^6$~K. Consequently, we get $T_i=2.469\times
10^6$~K and $T_e=2.475\times 10^6$~K, respectively (note that
$\Delta T=T_i-T_s\approx T_e-T_s\sim 3\times 10^4$~ K). Comparing
these critical temperatures with Eqs.~(\ref{teia}), (\ref{teijot})
and (\ref{sigmate}) we obtain $B_s=1.4\times 10^{13}$~G,
$1.5\times 10^{15}$~G and $5\times 10^{13}$~G, respectively. Thus,
again the calculations of the cohesive energy by Abrahams \&
Shapiro (\cite{as91}) represented by Eq.~(\ref{teia}) seem to be
the most adequate, while the calculations of Jones (\cite{j86})
lead to $B_s>10^{15}$~G. Also the electron case with $B_s=5\times
10^{13}~{\rm G}>B_q$ can be excluded in this case.

The observed time variability of the apparent drift rate,
including the change of the apparent drift direction in PSR
B0826-34 can be explained by time variations of the shielding
factor $\eta$ (Eqs.~(\ref{eta}), (\ref{eta2}), (A.7) and (A.8)),
which, according to Eq.~(\ref{eta1}), implies a time variability
of the surface temperature $T_s$. The detailed analysis of
sequences of drifting subpulses in PSR 0826-34 performed by Gupta
et al. (\cite{getal03}) indicates the following variations over
about 100 pulse sequences: the azimuth drift rate $D_r$ varies
from $25^\circ/P_1$ to $27^\circ/P_1$, drift velocity $v_d$ from
7.5 m/s to 9 m/s, $P_3$ from 1.03$P_1$ to 0.95$P_1$ and
$\hat{P}_3$ from 14.4$P_1$ to 13.3$P_1$. The change of the
apparent drift direction due to aliasing occurs when $P_3=1.0P_1$
and thus $\hat{P}_3=14P_1$. These approximately 100 pulse cycles
repeat in a quasi-periodic manner. During each cycle the shielding
factor increases by a few percent (around $\eta=0.36$) and the
surface temperature drops by about 2000 K (around $T_s=2.47\times
10^6$~K). Notice that similar (though erratic) variations of these
parameters, with an amplitude of about a few percent, were
observed in PSR 0943+10 (see footnote 3).

\subsection{PSR 2303+30}

GS00 reproduced the drifting subpulses in this pulsar (see their
Fig.~3) and argued that in this case $r_p/h\sim 5$, $N=12$ and
$\hat{P}_3\approx 23P_1$, which is consistent with $P_3\sim
1.9P_1$ obtained by Sieber \& Oster (\cite{so75}). For the
periphery of the polar cap we get $\eta=0.16$, $T_s=1.854\times
10^6$~K, $T_i=1.864\times 10^6$~K and $T_e=1.865\times 10^6$~K.
Comparison with Eqs.~(\ref{teia}), (\ref{teijot}) and (\ref{tee})
gives $B_s=9.6\times 10^{12}$~G (OK), $B_s=8.7\times 10^{13}$~G
(too high) and $B_s=3.7\times 10^{13}$~G (probably too high),
respectively. Thus again, the calculations of the cohesive energy
by Abrahams \& Shapiro (1991) seem to be most adequate.

This pulsar also shows an apparent change of the subpulse  drift
direction (see Fig.~ 3 in GS00), which can be explained by small
changes of: the azimuth drift-rate from $14.^\circ 4/P_1$ to
$15.^\circ 5/P_1$, the drift velocity $v_d$ from $\sim 10$~m/s to
$\sim 11$~m/s, $P_3$ from 2.1$P_1$ to $1.9P_1$ and $\hat{P}_3$
from $25P_1$ to $23P_1$ (notice, that similar variations of these
parameters were observed in PSRs 0826-34 and 0943+10). The
corresponding change of the shielding factor $\eta$ is about one
percent, which implies the change of the surface temperature $T_s$
of about 1000K (around $T_s=1.85\times 10^6$~K).

\subsection{PSR 2319+60}

GS00 reproduced the drifting subpulses in this pulsar (see their
Fig.~4) and argued that in this case $N=9$, $\hat{P}_3\sim 70P_1$,
$P_3\sim 7.8P_1$ and $r_p/h\sim 5$. For sparks drifting at the
periphery of the polar cap we get $\eta\sim 0.071$,
$T_s=1.236\times 10^6$~K, $T_i=1.239\times 10^6$~K and
$T_e=1.239\times 10^6$. Comparison with Eqs.~(\ref{teia}),
(\ref{teijot}) and (\ref{tee}) gives $B_s\sim 5\times 10^{12}$~G
(this value is lower than $B_d=8\times 10^{12}$~G and should be
rejected), $B_s=5\times 10^{13}$~G (this value is higher than the
photon splitting limit $B_q$ and should also be rejected) and
$B_s=1.36\times 10^{13}$~G. Thus, only the latter case seams
acceptable and therefore PSR 2319+60 should be considered as the
pulsar with a negatively charged polar cap (antipulsar in RS
terminology).

\subsection{PSR 0031-07}

GS00 reproduced drifting subpulses in this pulsar (see their
Fig.~5) and argued that in this case $N\geq 5$, $\hat{P}_3>34P_1$,
$P_3\sim 6.8P_1$ and $r_p/h\sim 4$. For the peripheral sparks we
get $\eta>0.089$, $T_s=1.551\times 10^6$~K, $T_i=1.556\times
10^6$~K and $T_e=1.557\times 10^6$~K. Comparison with
Eqs.~(\ref{teia}), (\ref{teijot}) and (\ref{tee}) gives
$B_s=7.2\times 10^{12}$~G (OK), $B_s=7.9\times 10^{13}$~G (too
high) and $B_s=2.2\times 10^{13}$~G (probably too high). Thus
again, the calculations of cohesive energy of Abrahams \& Shapiro
(1991) seem to be most adequate and calculations by Jones (1986)
are not suitable within our model.

\section{Discussion and conclusions}

Recent progress in the analysis of drifting subpulses in a number
of pulsars (DR99, DR01, Asgekar \& Deshpande 2001, van Leeuwen et
al. 2002, GMM01b, Gil \& Melikidze 2002 and Gil \& Sendyk 2003)
provided a strong support to the canonical RS pulsar model, which
relates the spark plasma filaments with subpulse-associated
subbeams of radio emission. The observed subpulse drift is
naturally explained by the inevitable ${\bf E}\times{\bf B}$
drift of these plasma filaments, at least qualitatively. However,
on the quantitative level, the ${\bf E}\times{\bf B}$ drift in a
pure vacuum gap is too fast as compared with observations.
 In this paper, we consider a general concept of the
charge depletion region (rather than a pure vacuum gap), with a
thermal outflow of $^{56}_{26}$Fe ions, coexisting with the
generation of $e^-e^+$ plasmas (CR80). The presence of an ion flow
decreases the accelerating gap potential drop and, in consequence,
the ${\bf E}\times{\bf B}$ drift-rate, as well as the amount of
surface heating due to back-flowing relativistic electrons
(positrons). As originally demonstrated by CR80, the extreme
sensitivity of the potential drop to the surface temperature makes
the gap thermostatically self-regulating, whenever there is both
an ion outflow and $e^-e^+$ discharge plasma. We argued that when
heating and cooling are nearly in balance, the surface temperature
$T_s$ oscillates within a very narrow range around its
quasi-equilibrium value, slightly below the critical temperature
$T_{i}$ ($T_e$) above which the ion outflow reaches the
co-rotational GJ density.

Our general model of the charge depleted acceleration region can
be applied not only to pulsars with positively charged polar caps
$({\bf\Omega}\cdot{\bf B}<0)$, but also to pulsars with negatively
charged polar caps $({\bf\Omega}\cdot{\bf B}>0)$, which are
usually interpreted within the so-called stationary flow models
(e.g. Arons 1981 and references therein) in which electrons flow
freely from the polar cap surface at the GJ charge density. Within
the model of a partially shielded acceleration region the
$^{56}_{26}$Fe ions are only marginally bound at a temperature
$T_s\lesssim T_i$, but, nevertheless, the non-stationary sparking
scenario can be realized in a way similar to that envisioned by RS
in their pure vacuum gap model (with strong binding assumed). We
suggest that such marginally bound ions are not much different
from marginally bound electrons (at $T_s\sim T_e$). In fact,
RS-type non-stationarity (sparks) can also arise if the thermionic
electron flow from the surface is not exactly at the GJ
corotational charge density. As we demonstrated, even small
departures from GJ charge density at the surface temperature
$T_s\approx T_e$ can result in the sparking discharge of the
shielded accelerating potential drop above the polar cap. Thus we
suggest that both pulsars with positively and negatively charged
polar caps can develop non-stationary inner acceleration regions,
similar to those invoked by RS. In consequence, the
spark-associated coherent radio emission due to instabilities in
the non-stationary $e^-e^+$ secondary plasma (e.g. Usov 1987,
Asseo \& Melikidze 1991, Melikidze et al. 2000) may originate
similarly in both pulsars $({\bf\Omega}\cdot{\bf B}<0)$ and
``antipulsars'' $({\bf\Omega}\cdot{\bf B}>0)$.

We estimated values of the polar cap surface temperature for six
pulsars (Table 1). The estimated values of $T_s$ lie in the range
between $\sim 1.0$ to $\sim 2.5$ million K. It is worth noting
that in a pure vacuum gap (in RS with $\eta=1$), the values of
$T_s$ would be higher by a factor of about 2, thus in some pulsars
from our list exceeding $4\times 10^6$~K at the polar cap. Present
and future x-ray satellite spectral measurements should be able to
constrain maximum surface temperatures of the polar cap heated by
intense spark discharges. One should also mention that our
estimates of $T_s$ were obtained under canonical assumptions that
${\cal R}_6=1$ (e.g. RS). However, if the radius of curvature of
surface magnetic field lines is much smaller than $10^6$~cm, then
the values of $T_s$ can be lowe even by a factor $2-3$. Thus,
our estimates of $T_s$ presented in Table 1 represent the upper
limits.

Our method is capable of determing both the actual surface
temperature $T_s$ and critical temperature $T_i$ or $T_e$, above
which the thermionic emission of iron ions or electrons reaches
the corotational GJ value. We found that radio pulsars operate at
$T_s$ approximately $10^4$~K lower than $T_i$ or $T_e$. Comparing
our observationally deduced values of critical temperatures with
estimates based on the calculations of the cohesive energy and/or
work function, we found the required values of the surface
magnetic field $B_s=bB_d$ (Table 1). Within our model we can
exclude the calculations of the cohesive energy by Jones
(\cite{j86}), represented in our paper by Eq.~(\ref{teijot}). In
most cases calculations of Abrahams \& Shapiro (\cite{as91})
represented by Eq.~(\ref{teia}) give suitable results, except PSR
0031-07 which is the best candidate for drifting subpulse pulsars
with a negatively charged polar cap. Also PSR 0869+74 seems to
belong to this category. Interestingly, these two pulsars require
the lowest shielding factors $\eta$ (Table 1) to fit the observed
and the theoretical drift-rates.

The sparks rotate due to the ${\bf E}\times{\bf B}$ drift in the
same direction as the neutron star rotates (lagging behind the
polar cap surface). As a matter of fact, the particle drift in the
inertial frame cannot exceed the stellar rotation speed within any
reasonable model. Thus, the observed direction of the subpulse
drift is fixed for a given pulsar and depends only on whether the
line-of-sight trajectory passes poleward or equatorward of the
magnetic axis. However, two pulsars from our sample demonstrate an
apparent change of the drift direction. We argued that
sense-reversals can be explained by small changes of the
drift-rate causing the $P_3$ value to fluctuate around the
relevant Nyquist boundary: $P_3\approx 1P_1$ in the case of PSR
0826-34 and $P_3\sim 2P_1$ in case of PSR 2303+30, corresponding
to fluctuation frequencies $f_3$ equal to 1 cycle/$P_1$ and 0.5
cycle$/P_1$, respectively. We believe that these changes are due
to small variations of the polar cap surface temperature. PSR
0540+23 probably also belongs to the category of pulsars with
$P_3\approx 1P_1$, showing apparent reversals of the sense of
subpulse drift (Nowakowski 1991), although its drifting patterns
are much more chaotic than in PSR 0826-34.

Up to now, it was believed that in conal profiles (Rankin 1983)
the values of $P_3$ periods covered the range from about 2 to
about 15 pulsar periods $P_1$ (corresponding to fluctuation
frequencies between about $0.5/P_1$ (Nyquist frequency) to about
$0.07/P_1$). It is well illustrated in Fig.~4 and Table~2 in
Rankin (1986). In this paper we broadened this range by adding at
least one pulsar PSR 0826+23 with $P_3\sim P_1$ (Table~1), and
probably PSR 0540+23 (Nowakowski 1991) also belongs to this
category. Perhaps there are many more pulsars with a fast drift
corresponding to $P_3$ values well below ``Nyquist limit''.

It is worth emphasizing again that the subpulse drift is widely
considered as a crucial clue towards solving the long standing
mystery of pulsar radio emission. Despite its potential
importance, this phenomenon has not attracted much theoretical
attention beyond that of RS model. A simple explanation of this
fact is that probably it is very difficult to propose a theory
that would be as natural as the RS model, at least qualitatively.
Moreover, this is the only model offering quantitative predictions
that can be compared with observations. We review in this paper a
number of problems appearing when the existing theory is confronted
with the current pulsar data and argue that all of them disappear
when the original RS model is modified to include a thermionic
outflow of ions or electrons, coexisting with the
electron-positron plasma produced in spark discharges. We also
emphasize the importance of the strong non-dipolar surface
magnetic field driving the spark avalanches, as well as resonant
inverse Compton scattering as the mechanism providing seed photons
for these avalanches.

Finally, we believe that the phenomenon of drifting subpulses is a
manifestation of a more general phenomena related to sparking
discharges of the ultra-high potential drop above the polar cap of
radio pulsars. Therefore, the result of this paper strongly
suggest that spark-associated plasma instabilities (Usov 1987,
Asseo \& Melikidze 1998) play an important role in generation of the
observed coherent pulsar radio emission (e.g. Melikidze et al.
2000).

\begin{acknowledgements}
 This paper is supported in part by the KBN grant 2
P03D 008 19 of the Polish State Committee for Scientific Research.
JAG acknowledges the renewal of the Alexander von Humboldt
Fellowship and thanks G. R\"{u}diger for hospitality at the
Astrophysikalisches Institut Potsdam, where this work was done. We
also thank E. Gil and U. Maciejewska for technical help.
\end{acknowledgements}

\appendix
\section{Inner acceleration region}

\subsection{Charge depleted acceleration region above the polar
cap}

The acceleration potential drop above the polar cap results from
the deviation of a local charge density $\rho$ from the
corotational GJ density \be\rho_{GJ}=-\frac{{\bf\Omega}\cdot{\bf
B_s}}{2\pi c} , \label{rogj}\ee where ${\bf\Omega}$ is the pulsar
spin vector and ${\bf B}(r)$ is the pulsar magnetic field, which
should be purely dipolar at altitudes of a few stellar radii
$R_*=10^6$~cm, where the pulsar radio emission is expected to
originate (e.g. Kijak \& Gil 1998 and references therein), but
could be highly nondipolar (i.e. having a pronounced small-scale
spatial structure) at the polar cap surface. For reasons of
generality we describe the surface magnetic field by \be B_s=bB_d
, \label{bees}\ee where \be B_d=2\times
10^{12}(P_1\dot{P}_{-15})^{0.5}~{\rm G} \label{bede} \ee is the
dipolar component of the pulsar magnetic field at the pole, $P$ is
the pulsar period in seconds and $\dot{P}_{-15}=\dot{P}/10^{-15}$
and $b>1$. This defines the actual radius of the polar cap \be
r_p=b^{-0.5}10^4P_1^{-0.5}~{\rm cm} ,\label{erpe} \ee (e.g. GS00).
We also introduce the curvature radius of nondipolar surface
magnetic field lines ${\cal R}_6={\cal R}/R_*\lesssim 1$, where
$R_*=10^6$~cm is the neutron star radius (note that for a purely
dipolar field $R_6\sim 300$ near the polar cap). Within our model
of the charge depleted region the local charge density
$\rho=(1-\eta)\rho_{GJ}$, where $\eta=1-\rho_i/\rho_{GJ}$ (or
$\eta=1-\rho_e/\rho_{GJ}$) and $\rho_i$ (or $\rho_e$) is the
charge density of outflowing thermionic ions (or electrons).
Within the acceleration region the potential $V$ and the electric
field $E$ (parallel to the magnetic field ${\bf B_s}$) are
determined by one-dimensional Poisson equation
$d^2V/dz^2=-4\pi(\rho-\rho_{GJ})=4\pi\eta\rho_{GJ}$, with the
boundary condition $dV/dz=0$ at $z=h=h_{max}$ and $dV/dz=E_{max}$
at $z=0$ (see below for determination of $h$). Since
$\rho_{GJ}\approx\pm B_s/cP$, where the sign $+(-)$ corresponds to
the ion (electron) flow, the solution of Poisson equation gives
$E(z)=\pm\eta(4\pi/cP_1)B_s(h-z)$ and $\Delta
V=\int_0^hE(z)dz=\pm\eta(2\pi/cP_1)B_sh^2$ or \be \Delta
V=\pm\eta\Delta V_{max} ,\ee where \be \Delta
V_{max}=\frac{2\pi}{cP_1}B_sh^2 \ee is the acceleration potential
drop within a pure vacuum gap (RS) corresponding to $\eta=1$
($\rho_i=0$ or $\rho_e=0$).

The height $h$ of the acceleration region with ultra-high
potential drop $\Delta V$ described by Eqs.~(A.1) and (A.2) is
determined by a quasi-steady discharge via magnetic conversion of
high energy $\hbar\omega>2mc^2$ photons into $e^-e^+$ pairs that
develop cascading sparks. Thus $h=l_{ph}$, where $l_{ph}$ is the
mean free path of an energetic photon to produce a pair. At least
several possible models of the inner acceleration region were
considered by different authors (see Zhang et al. 2000; ZHM00
hereafter and GM02 for review). Among them, the most promising one
with respect to the possibility of creating an effective
acceleration region (``vacuum gap'') seems to be the
Near-Threshold $(\hbar\omega=2mc^2{\cal R}/l_{ph})$ model,
involving the resonant ICS seed photons $(\hbar\omega=2\gamma\hbar
eB_s/mc)$ in the strong magnetic field $B_s>0.1B_q\sim 5\times
10^{12}$~G (see GM01 and GM02 for details of the so-called
ICS-NTVG model). Unlike in the RS case with curvature seed
photons, the mean free path of ICS seed photons in ICS-NTVG
model is approximately equal to the mean electron path
$l_e=0.0027\gamma^2B_{12}^{-1}T_6^{-1}$ to emit an energetic
photon with energy $\hbar\omega=2\gamma\hbar eB/mc$ exceeding a
pair formation threshold (ZHM00, GM01, GM02), where
$B_{12}=B_s/10^{12}$~G. Since $2mc^2{\cal R}/h=2\gamma\hbar
eB_s/mc$, then the minimum Lorentz factors $\gamma$ required for
the ICS dominated pair production under the Near Threshold
conditions is \be \gamma_{min}=(2.18\times 10^7)b^{-1}{\cal
R}_6h^{-1}(P\dot{P}_{-15})^{-0.5} ,\ee (see Eq.~(2) and GM01 and
references therein for details). Now using $h_{ICS}=h=l_e$ with
$\gamma=\gamma_{min}$ we can obtain the expression for the height
of the ICS dominated acceleration region with a strong magnetic
field \be h_{ICS}=(8.6\times 10^3){\cal
R}_6^{2/3}b^{-1}(P\cdot\dot{P}_{-15})^{-1/2}T_6^{-1/3}~{\rm
cm}\label{haics}\ee and thus $\gamma_{min}\approx 2.5\times
10^3T_6^{1/3}{\cal R}_6^{1/3}$, which reproduces Eq.~(10) in GM02
\be h_{ICS}=(5\times 10^3){\cal
R}_6^{0.57}b^{-1}P^{-0.36}\dot{P}_{-15}^{0.5}~{\rm
cm},\label{haics1}\ee if $T_6=T_s/10^6$~K is expressed by their
Eq.~(12) and $\eta=\xi=k=1$. According to Eqs.~(2), (A.1) and
(A.2), the acceleration potential drop is \be \Delta V=(8\times
10^{12})\eta b^{-1}{\cal
R}_6^{4/3}T_6^{-2/3}P^{-3/2}\dot{P}_{-15}^{-1/2}~V, \ee which
reproduces Eq.~(11) in GM02 \be \Delta V_{ICS}=(5.2\times
10^{12}){\cal
R}_6^{1.14}b^{-1}P^{-1.22}\dot{P}_{-15}^{-0.5}~V,\nonumber\ee
again under conditions mentioned above. The electrons (positrons)
can be accelerated by this potential drop to the Lorentz factor
values \begin{eqnarray} \gamma_{acc}= \frac{e\Delta V}{mc^2}= \nonumber \\
1.8\times 10^7\eta b^{-1}{\cal
R}_6^{4/3}P^{-3/2}\dot{P}_{-15}^{-1/2}T_6^{-2/3} .\end{eqnarray}

It should be noted here that in the case of positively charged
polar cap the full accelerating potential drop $\Delta
V=\eta\Delta V_{max}+(1-\eta)\Delta V_{SCLF}$, where $\Delta
V_{SCLF}\approx 3\times 10^{11}P_1^{-1/2}(B_s/10^{12}{\mathrm
G})^{1/2}(h/10^4~{\mathrm{cm}})$~V is the potential drop arising
due to ions inertia in the space-charge-limited flow (e.g. CR80).
If $h\sim h_{ICS}<10^3$~cm (Eq.~(A.9)) and $\Delta
V_{max}\sim\Delta V_{ICS}$ (Eq.~(A.11)) then for $\eta\sim 0.1$
(Table 1) $\Delta V\sim\eta\Delta V_{max}$. Thus, within the
acceleration region the SCLF potential drop can be neglected, even
when the general relativistic (GR) effect of inertial frame
dragging (Muslimov \& Tsygan 1992) is taken into account. In fact,
the GR-induced electric field $E_{GR}$ grows quasi-exponentially
from zero at the surface to the maximum value at the height
approximately equal to the polar cap radius $r_p\gg h<10^3$~cm.
Thus, the value of the integral $\int_0^hE_{GR}dz$ must be
negliglible compared with $\Delta V$ (e.g. Eq.~(A.10)).

\subsection{${\bf E}\times{\bf B}$ drift in the acceleration
region}

If the actual charge density $\rho<\rho_{GJ}$ (thus $\eta<1$),
then the plasma within the acceleration region does not corotate
with the neutrons star. This is the so-called ${\bf E}\times{\bf
B}$ drift, which results in a slow motion of $e^-e^+$ plasma in
the direction perpendicular to the planes of the local surface
magnetic field lines. Effectively, any filamentary lateral
structures (sparks) circulate slowly around the local magnetic
pole with a velocity $v_d=c\Delta E_\perp/B_s$, where $\Delta
E_\perp$ is the component of the electric field caused by charge
depletion $\Delta\rho=\rho_{GJ}-\rho=\eta\rho_{GJ}$, $B_s$ is the
local surface magnetic field and $c$ is the speed of light.
According to Eq.~(30) in RS, this electric field $\Delta
E_\perp=\Delta V/(r_p-d)$, where $\Delta V$ is the potential drop
described by Eqs.~(A.1) and (A.2), $r_p$ is the radius of the
polar cap (Eq.~(\ref{erpe})), and $d<r_p$ is the circulation
distance of sparks from the local magnetic pole.

Each spark completes one full circulation around the pole in a
time interval $\hat{P}_3\approx 2\pi d/v_d$ seconds. Using
Eqs.~(A.5) and (A.6) we obtain the so-called tertiary periodicity
(expressed in pulsars periods $P_1$) \be
\hat{P}_3=\frac{P_1}{\eta}\left(\frac{r_p}{h}\right)^2s(1-s)=
\frac{P_1}{\eta}\frac{d}{h}\left(\frac{r_p-d}{h}\right) ,
\label{pe3}\ee where $s=d/r_p$. Consequently, the azimuthal
drift-rate $D_r=d\chi/dt$ (where $\chi$ is the magnetic azimuth
angle of the circulating spark) is \be
D_r=\frac{360^\circ}{\hat{P}_3}=\frac{360^\circ}{P_1}\frac{\eta
h^2}{d(r_p-d)} .\ee We assume reasonably that the biggest
contribution to the ${\bf E}\times{\bf B}$ drift occurs at the
beginning of the spark discharge, when $\Delta V\sim\eta\Delta
V_{max}$ (Eqs. (A.5) and (A.6)) that is when the screening due to
exponentially growing $e^-e^+$ pair plasma is weak. This justifies
the form of our Eqs.~(A.13) and (A.14).

Assuming $d=r_p/2$ (thus $s=0.5$), we have
$\hat{P}_3=0.25(r_p/h)^2$, which for $\eta=1$ (a pure vacuum gap)
reproduces Eq.~(32) in RS to within a factor of 0.25. This
discrepancy follows from the factor 2 obviously missing in their
Eq.~(31). In fact, RS used the approximation $<ae>\sim r_p/2$
expressed explicitely just above their Eq.~(31), in which,
however, they put $d=<ae>=r_p$, inconsistent with their
assumption.

As originally argued by RS, the spark model very strongly suggests
that \be \hat{P}_3=NP_3 ,\ee where $N$ is the number of
circulating sparks, and $P_3=1/f_3$ is the basic drifting
periodicity determined either visually or measured from the
dominant feature $f_3$ in the fluctuation spectrum (e.g. Backer
1973). If necessary, the value of $P_3$ has to be corrected for
aliasing to allow to describe adequately the relationship between
the tertiary periodicity $\hat{P}_3$ and the number of circulating
sparks $N$.

\section{Heat transport at the polar cap surface}

\subsection{Polar cap physics}
An inherent property of the polar cap is the presence of the
ultra-strong magnetic field (ignored by CR80), which is almost
perpendicular to the surface. The presence of such strong magnetic
field affects the equation of state, as well as the transport
processes in the outermost surface layers of the neutron star. The
state of matter and the heat transport regime in the polar cap
region is determined by a number of quantities described in the
following subsections B.1.1. -- B.1.8:

\subsubsection{Surface density}
The existence of a magnetic field in the order of magnitude of
$\sim 10^{13}$G enhances the binding energy of the electrons. This
changes drastically the density profile in the very surface
layers, the pressure (in the Thomas--Fermi approximation) vanishes
at the so--called zero pressure condensation density $\rho_s$
which may be regarded as the surface density for a 3--dim $bcc$
crystal (R\"ognvaldsson et al. 1993). The most recent result for
that density is obtained by Lai (2001) \be \rho_{\mathrm s}\simeq
561 AZ^{-3/5}B_{12}^{6/5} ~ {\rm g\; cm}^{-3} . \ee Here we
consider the neutron star surface to be composed of iron, i.e.
$A=56$ and $Z=26$. Thus, the surface matter density is
$\rho_{\mathrm s}(B_{12}=1)= 4.45\times 10^3$g cm$^{-3}$ and
$\rho_{\mathrm s}(B_{12}=10)= 7.05\times 10^4$g cm$^{-3}$,
respectively.

\subsubsection{State of aggregation}
The state of aggregation of the polar cap matter is determined by
the Coulomb parameter $\Gamma= (Ze)^2/(akT)$, which measures the
ratio of electrostatic and thermal energy; $a$ is the
Wigner--Seitz cell radius $\propto \rho^{-1/3}$. Once the Coulomb
parameter exceeds $\Gamma = 170$, the ions form a crystal (see
e.g. Slattery et al. 1980), the corresponding melting density is
\be \rho_{\mathrm m} =\left(\frac{170\,kT}{(Ze)^2}\right)^{3}
        \frac{3\,A\,m_{\mathrm u}}{4\pi}\;= 76\;T_6^3\;{\rm g
        cm}^{-3},
\ee where $m_{\mathrm u}$ is the atomic mass unit. Since
$\rho_{\mathrm m} \ll \rho_{\mathrm s}(B_{12}=1)$ up to
temperature $T_6 = 3$,  the polar cap matter will not be melted
but consist of crystallized iron.

\subsubsection{Degree of ionization}
For $\rho \ge 10^5$g cm$^{-3}$ the atoms are fully ionized due to
the pressure independently of the temperature, because the mean
volume of a free electron becomes smaller than the volume of the
$1s-$ orbitals. For $\rho_{\mathrm s}(B_{12}=1$ or $10)$ the mean
ionization per iron atom in the temperature range $T_6 \ge 1$ is
$24$ or $25$, respectively (Schaaf 1988), i.e. the atoms in the
polar cap are almost completely ionized. The iron ions are
non--degenerated and not affected by the magnetic field, i.e. the
phonon spectrum of the crystal remains the same as for the zero
magnetic field $(B=0)$ case.

\subsubsection{Degree of degeneracy}
The Fermi temperature for the iron polar cap is given by  \be
T_{\mathrm F}= 6\times
10^9\left(\sqrt{1+0.6\rho_6^{2/3}}-1\right)\;{\rm K}, \ee (see
e.g. Hernquist 1984). In the presence of a quantizing magnetic
field the electron chemical potential  is reduced. If only the
lowest Landau level is populated, the Fermi temperature is \be
T_{\mathrm F}= 6\times
10^9\left(\sqrt{1+200\rho_6^2/B_{12}^2}-1\right) \;{\rm K}. \ee
Thus, for the typical densities and magnetic field strengths in
the polar cap region the electrons there are in the state of a
complete degeneracy, provided that the surface is not hotter than
a few times $10^6$~K.

\subsubsection{Degree of being relativistic}
That state of the electron gas is determined by the relativistic
parameter $x=p_{\mathrm F}(B=0)/m_{\mathrm e}\;c$ (Yakovlev \&
Kaminker 1994). The Fermi momentum at $B=0$ is given by \be
p_{\mathrm F}(B=0)= \hbar\left(3\pi^2\frac{Z\;\rho}{A\;m_{\mathrm
u}}\right)^{1/3}\;. \ee For $\rho_{\mathrm s}(B_{12}=10)$ we find
$p_{\mathrm F} \approx 8.8\times 10^{-18}$g cm s$^{-1}$ which is
much less than $m_{\mathrm e}\;c$. Therefore, the electrons in the
polar cap region are non--relativistic.

\subsubsection{Number of populated Landau levels}

As shown by  Yakovlev \& Kaminker (1994) for conditions realized
at the polar cap, the strong magnetic field $> 10^{12}$G has a
quantizing effect on the electron motion. How many Landau levels
are typically populated? The maximum number of these levels is
given by (see e.g. Fushiki et al. 1989, Hernquist 1984) \be
n_{\mathrm max} = \frac{p_{\mathrm F}^2(B=0)\;c}{2\;e\;\hbar\;B}
\approx 14\frac{\rho_6^{2/3}}{B_{12}} . \ee For $B_{12} \approx
10$ and $\rho \le \rho_{\mathrm s}(B_{12}=10)= 7\times 10^4$g
cm$^{-3}$, only the lowest Landau level ($n=0$) is populated.

\subsubsection{Debye temperature}

For temperatures larger than the Debye temperature
$\Theta_{\mathrm D}$ the specific heat of the $bcc$ crystal
lattice can be approximated by its classical value
$C_{\mathrm{lattice}} = 3 k n_{\mathrm i}$, where $n_{\mathrm i}$
is the number density of ions. According to Yakovlev \& Urpin
(1980), the Debye temperature as a function of the ion plasma
frequency $\omega_{\mathrm p}$ is given by \be \Theta_{\mathrm D}=
0.45\frac{\hbar\;\omega_{\mathrm p}}{k} \approx 1.6\times
10^6\;\rho_6^{1/2}\;{\rm K}. \ee The presence of a strong magnetic
field does not change $\Theta_{\mathrm D}$ drastically. Since
$\rho_{\mathrm s}(B_{12}=10) \ll \rho_6$, we find for the polar
cap layer $T > \Theta_{\mathrm D}$ and the specific heat of the
crystal there can be regarded as caused by classical lattice
vibrations.

\subsubsection{Depth of heat deposition}

The depth of the heat deposition due to the bombardment of the
polar cap by ultrarelativistic electrons/positrons is measured in
so--called radiation lengths (see CR80). For $^{56}_{26}$Fe ions
the radiation length $y= 14$~g cm$^{-2}$. Therefore the
corresponding depth $L \approx y/\rho_{\mathrm s}$, and for
$B_{12}=10$ we have $L\sim 2\times 10^{-4}$~cm.

\subsection{Cooling timescales}

The polar cap surface is heated by the bombardment of
back--flowing relativistic electrons produced in the pair
producing spark discharge of the accelerating potential drop. Once
the total charge density $\rho_t=\rho_{i,e}+\rho_\pm$ approaches
the GJ value, the intense heating ceases until the next spark
recurs. In the absence of heating the region beneath the spark
cools down rapidly. The cooling process was considered by CR80,
who found the characteristic cooling time $\tau_c=0.32
(T_s/10^6~K)^{-6}(\rho_s^m/10^5{\rm g cm}^{-3})^{7/3}$ seconds,
where $T_s$ is the actual surface temperature and $\rho_s^m$ is
the actual matter density just beneath the polar cap surface.
Adopting $T_s=10^6$ K and $\rho_s^m=10^5$~g cm$^{-3}$ CR80
obtained a small fraction of a second for $\tau_c$ (which could be
a very tiny fraction of a second if $T_s$ is slightly higher and
$\rho_s^m$ is lower). It seems therefore appropriate to reconsider
their derivation, since much more precise estimates for the
transport coefficients and their magnetic field dependence in the
uppermost surface layer are currently available. In what follows
we omit the superscript $m$ denoting the matter density $\rho^m$.

As we have shown in Appendix B.1. the magnetized ($B_{12}=10$)
surface of the polar cap in isolated (non--accreting) neutron
stars has a density $\rho_s \approx 7\times 10^4$ g cm$^{-3}$ and
at temperatures of $T_s \ge 10^6$ K consists of almost completely
ionized iron ions which form a solid $bcc$ crystal lattice having
a Debye temperture well below $10^6$ K. The electrons in that
lattice form a degenerated non--relativistic gas which populates
the ground Landau level only.

Let us now estimate the characteristic cooling time for the
strongly magnetized outermost surface layer of the  polar cap. The
cooling time can be deduced from the heat transport equation. The
heat transport in the polar cap region can be considered in a very
good approximation as purely parallel to the magnetic field lines,
which are almost perpendicular to the surface. Therefore, the heat
transport is well described by a one--dimensional heat transport
equation with the use of the time coordinate $t$ and the spatial
length coordinate $l$ (parallel to the magnetic field lines).
Without sinks and sources of energy and together with the boundary
condition that the heat is radiated away from the very surface
according to $-\kappa\;\frac{\partial T}{\partial l} =
\sigma\;T^4$ (where $\sigma$ is the Stefan-Boltzman constant), the
heat transport equation reads \be C\frac{\partial T}{\partial t} =
\frac{\partial}{\partial l}\;\left( \kappa\;\frac{\partial
T}{\partial l}\right)\;. \ee By assuming in the very thin surface
layer an almost uniform heat conductivity, and using ${\partial
T}/{\partial t} \approx {T}/{\tau_{\mathrm c}}$ and ${{\partial}^2
T}/{\partial l^2} \approx {T}/{L^2}$, we
 obtain the e-folding cooling time
\be \tau_{\mathrm cool} =\left(\frac{L^2 C}{\kappa}\right)\;, \ee
where $L$ is the depth of heat deposition, $\kappa$ is the thermal
conductivity and $C$ is the specific heat per unit volume.

Heat can be stored both in lattice vibrations and in the electron
gas. In the parameter range under consideration  these
contributions to the specific heat are comparable, thus \be C =
C_{\mathrm{lattice}} + C_{\mathrm{el}}\;. \ee For the number of
electrons per nucleon corresponding to $^{56}_{26}$Fe ions, the
contribution of the non--relativistic degenerated electron gas to
the specific heat is \be C_{\mathrm{el}} \approx 1.05 \times
10^{11} \rho_6^{1/3}\;T_6\ {\rm erg\;K}^{-1}{\rm cm}^{-3} , \ee
(see Landau \& Lifshitz 1969). Since at the polar cap $ T
> \Theta_{\mathrm D}$ the lattice specific heat is given by
\begin{eqnarray} C_{\mathrm{lattice}} =
3\;n_i\;k=\frac{3k}{Am_u}\rho\approx \nonumber \\ 4.4\times
10^{12} \rho_6\ {\rm erg\;K}^{-1}{\rm cm}^{-3}\;. \end{eqnarray}
Thus, we obtain for the specific heat per unit volume at the polar
cap region \be C \approx 4.4 \times 10^{12}\rho_6\left(1 +
0.024\rho_6^{-2/3}T_6\right) {\rm erg\;K}^{-1}{\rm cm}^{-3} . \ee

For estimates of the thermal conductivity we rely on the results
obtained by Schaaf (1988, 1990). He investigated the cooling of a
neutron star with magnetized envelopes and therefore calculated
carefully the transport coefficients in the density range $\rho
\le 10^{10}$ g cm$^{-3}$ for a temperature of about $10^6$ K and
magnetic field strengths of $10^{12}-10^{13}$ G. The electron--ion
collisions are insignificant as transport processes, because the
iron ions form a crystal lattice in the considered
temperature--density range. Since in the outermost layers of the
neutron star the heat is transported both by radiation and by
conduction, $\kappa = \kappa_{rad} + \kappa_{cond}$, we have to
estimate first the relative importance of these two cooling
mechanisms.

For densities $\rho \ge 10^3$ g cm$^{-3}$, the radiative transport
in the temperature and magnetic field range under consideration is
limited by the opacity due to free--free transitions,
$\tilde{\kappa} \approx \tilde{\kappa}_{ff}$ (see Fig. 3 of Schaaf
1990). For $\rho > \rho_{\mathrm s}(B_{12}=1)$ the component of
$\tilde{\kappa}_{ff}$ parallel to the magnetic field is about
$10^3$g$^{-1}$cm$^{2}$. The corresponding radiative heat
conductivity  $\kappa_{rad}=16 \sigma T^3/(3 \tilde\kappa \rho_s)
\approx 6.8\times 10^7~{\rm erg\ cm}^{-1}{\rm s}^{-1}{\rm
K}^{-1}$. This value varies not much with increasing $B_{12}$
because the growth of $\rho_{\mathrm s}$ compensates at least
partially the decrease of $\tilde{\kappa}$. This $\kappa_{rad}$
has to be compared with the heat conductivity limited by
electron--phonon collisions; the contribution of
electron--impurity collisions is negligible for such low
densities. For densities below $10^6$ g cm$^{-3}$ the parallel
component of the conductivity tensor depends strongly on the local
magnetic field strength because at low densities the quantization
effects  on the electron motion are dominant (Schaaf 1988, Fig
4.13); $\kappa_{cond}(\rho_{\mathrm s}(B_{12}=1)) \approx
1.6\times 10^{11}$ erg cm$^{-1}$ s$^{-1}$ K$^{-1}$ while
$\kappa_{cond}(\rho_{\mathrm s}(B_{12}=10)) \approx 1.6\times
10^{12}$ erg cm$^{-1}$ s$^{-1}$ K$^{-1}$. Therefore, the heat
transport in the polar cap region is nearly solely determined by
its electronic part, $\kappa \approx \kappa_{cond}$.

Now we can estimate the e-folding cooling time in the magnetized
outermost surface layer of the polar cap. Using B.1.8 for the
depth of heat deposition and assuming that the heat by the
bombardment of the polar cap surface is released in a depth of
about $10$ radiation length (see CR80) we obtain \begin{eqnarray}
\tau_{\mathrm cool}(B_{12}=1) \approx 200 \mu {\rm s} \;\; {\rm
and}\;\; \tau_{\mathrm cool}(B_{12}=10)\nonumber \\ \approx 1 \mu
{\rm s} \end{eqnarray} Therefore, the strong local surface
magnetic field is also the reason for an extremely short
characteristic cooling time in the polar cap surface layer. For
the thermostatic regulation described in this paper, the cooling
by a few percent of $10^6$~K may proceed in a time interval as
short as 10 ns. Note that the independent estimate of the cooling
time as given by CR80 is \be t_{cool}= \left(\frac{T-T_0}{\sigma
T_0^4}\right)^2 \kappa\; C , \ee which for values of $\kappa$ and
C given above, $B_{12}=10$ , $T_0=10^6$~K and $T-T_0=10^4$~K
yields a cooling time $t_{cool} \approx 20$~ns, a few percent of
our e-folding time $\tau_{cool}$.

\subsection{Heating timescales}

The inflow of energy per unit time from the bombardment of the
polar cap surface with ultrarelativistic electrons (positrons)
$\eta\gamma_{\rm acc}\;m_{\rm e}\;c^3\;n_{\rm GJ}$  must be
balanced by the change of the internal energy within the polar cap
volume $\propto r_{\rm p}^2\;L$, which is $CL\ \partial T/\partial
t$, where the Lorentz factor $\gamma_{\rm acc}$ is given by
Eq.~(A.13) and depends on the shielding factor $\eta$ as well as
on the local temperature, magnetic field strength and curvature.
The characteristic heating time can be defined by \be \tau_{heat}
= \frac{C\;L\;T}{\eta\;\gamma_{\rm acc}\; m_{\rm e}\;c^3\;n_{\rm
GJ}}\;\;. \ee We will give a lower limit for $ \tau_{heat}$, i.e.
consider the situation $\eta=1$, when the accelerating potential
drop is not yet shielded by thermionic ions (electrons). Using the
expressions derived in Appendix A.1. and B.1. we find the
characteristic heating time \begin{eqnarray} \tau_{heat} \approx
7.2\times 10^{-8}\rho_6\; T_6^{5/3}L_{-3}P_1^2{\cal R}_6^{-4/3}
\nonumber \\ \times\left(1 + 0.024\rho_6^{-2/3}T_6\right)\;\;{\rm
s}, \end{eqnarray} which returns for the $B_{12}=10$,
$\rho_6=0.07$, $L_{-3}=2$, $P_1=1$, ${\cal R}_6=0.1$ and $T_6=1$ a
characteristic heating time $\tau_{heat} \approx 250$ ns. Note
that actually $\tau_{heat} \propto \eta^{-2}$ and for pulsars
considered above $\eta < 0.2$. Therefore, the real characteristic
heating time should be of the order of 10~$\mu$s, which is about
$(10-40) h/c$, as determined by RS.

\section{Electron flow $({\bf\Omega}\cdot{\bf B}>0)$}

Akin to the case ${\bf\Omega}\cdot{\bf B}<0$, where positively
charged iron ions are pulled out from the surface to contribute to
the GJ density in the acceleration region, for the other ``half''
of pulsars, frequently called as ``antipulsars'', the situation
with ${\bf\Omega}\cdot{\bf B}>0$ has to be considered. The
question is: what is the threshold surface temperature $T_s$ (in
case of the thermionic particle extraction) or - electric field
$E_\|^s$ (in case of the field emission particle extraction),
below which the binding energy of the electrons in the polar cap
matter is significant to allow for the establishment of the charge
depleted acceleration region.

\subsection{Thermionic electron emission}

In order to screen totally the vacuum gap electric field $E_\|$
the thermionic extracted electrons have to provide the
Goldreich-Julian charge density. Since the electrons are nearly
instantenously accelerated to relativistic energies, the current
density due to this thermionic emission $j_{th}$ must somewhat
exceed the Goldreich-Julian current density, i.e. \be
j_{th}>n_{GJ} e\ c , \ee (see Usov \& Melrose 1995). The
thermionic current density is described by the Richardson-Dushman
equation (e.g. Gopal \cite{g74}) by \be
j_{th}=\frac{em_e}{2\pi^2\hbar^3}(kT)^2
exp\left\{-\frac{w}{kT}\right\} . \ee The work function $w$ has to
exceed somewhat the electron Fermi energy $\epsilon_F$ (Ziman
\cite{z74}). Since it is a very complicated task to calculate $w$
for the electrons at the neutron star surface, a task which has
not yet been solved, we will use $w\approx\epsilon_F$ as a
reliable lower limit.

In the outer layers of the polar gap where the matter density is
much lower than $10^6 {\rm g~cm}^{-3}$, and the temperature $\ll
10^7$~K, the electrons are degenerated but non-relativistic, i.e.
their Fermi energy $\epsilon_F$ is related to the Fermi-momentum
$P_F$ by \be \epsilon_F=\frac{P_F^2}{2m_e^*} , \ee where for
densities $<10^6 {\rm g~cm}^{-3}$ the effective electron mass
$m_e^*$ coincides almost exactly with its vacuum value $m_e$.
Under conditions of the magnetized polar cap, where the magnetic
field strength is of the order of $10^{13}$~G, up to densities of
$\sim 10^5$g cm$^{-3}$ only the ground Landau level $(n=0)$ is
populated. In that case the Fermi-momentum of the magnetized
electron gas is given by Fushiki et al. (\cite{fgp89}) \be
P_F(n=0)=\frac{2\pi\hbar^2n_ec}{eB} .\ee Correspondingly, the
electron Fermi-energy reads \be
\epsilon_F(n=0)=\frac{2\pi^4\hbar^4c^2Z^2\rho^2}{e^2m_u^2m_eA^2B^2}.
\ee Inserting $\rho=\rho_s(B)$ from Eq.~(B.1), the electron Fermi
energy at the solid iron surface of the polar cap is only a
function of the magnetic field \be \epsilon_F=1.58\times
10^{-9}B_{12}^{2/5} ~ {\rm erg} \approx 986 B_{12}^{2/5} {\rm eV}.
\ee Approximating $\epsilon_F\simeq w$ and with the
Goldreich-Julian charge number density $n_{GJ}={2\pi
B}/({P_1}{2\pi ce})={B}/({P_1ce})$, where $P_1$ denotes the
rotational period of the pulsar, we can rewrite Eq.~(C.1) in the
form \be 2.78\times 10^{-12}\frac{B_{12}}{P_1}\leq
T_6^2exp-\left\{11.45\frac{B_{12}^{2/5}}{T_6}\right\} , \ee where
$T_6=T_e/10^6$~K and $B_{12}=B_s/10^{12}$~G. This equation can be
solved iteratively for given values of $P_1$ and $B_{12}$. By
fitting dependencies on $P_1$ and $B_{12}$ we obtain an explicit
expression for critical electron temperature (Eq.~(C10)) in the
form \be T_e\simeq(4.5\times 10^5)B_{12}^{0.4}P_1^{-0.04} ~{\rm
K}.\label{cetee} \ee For comparison see Eq.~(10) in Usov \&
Melrose (1996) and Eqs.~(2.13) in Usov \& Melrose (1995), which we
reproduced and refined here. The numerical factor 4.5 takes into
account the most recent estimate of zero pressure surface matter
density obtained by Lai (\cite{l01}) for a 3 dimensional $bcc$
crystal (see Appendix B.1.).

\subsection{Field electron emission}

When the thermionic electron emission at the polar cap is not
efficient, perhaps because of the fact that the neutron stars have
been cooled down too much, still a field emission mechanism may
provide a sufficiently strong electron flow. A non-zero $E_\|$
will cause a quantum mechanical tunneling of the electrons through
the barrier provided by the work function and (e.g. Jessner et al.
\cite{jetal01}). The field emission current density of electrons
is given by \be j_{f}=ME_\| exp\{-N/E_\|\} ,\ee (e.g. Beskin
1982), where $M=3\times 10^{16}B_{12}w^{-1/2} {\rm s}^{-1}$ and
$N=2\times 10^{14}w^{3/2} {\rm Vm}^{-1}$ and $w$ measures the work
function in keV. Equating (as in the case of thermionic emission)
the Goldreich-Julian outflowing current density with $j_{f}$, and
approximating again $w\simeq \epsilon_F\simeq 986B_{12}^{2/5} {\rm
eV}$, the threshold electric field $E_\|^s$, above which the field
emission of electrons becomes effective is given by \be 1\leq
3\times 10^4P_1B_{12}^{-1/5}E_\| \{exp-(1.96\times
10^{14}B_{12}^{3/5}/E_\|)\} .\ee This equation can be solved
iteratively for given values of the pulsar period $P_1$ and the
magnetic field $B_{12}$. For $P_1=0.1$ and $B_{12}=1$ we find
$E_\|^s\gtrsim 2.05\times 10^{13} {\rm Vm}^{-1}$, a value
certainly too large to occur at the polar cap surface (see also
Appendix A in Gil \& Melikidze 2002).

 \clearpage
\rotatebox{90}{
\begin{tabular}{ccccccccccccccccc}
\multicolumn{6}{l}{\bf Table 1. Parameters of polar cap physics}&&&&&&&&&&\\
&&&&&&&&&&&&&&&& \\
  \hline\hline
 { }& {(1)}& {(2)}& {(3)}& {(4)}&
{(5)}& {(6)}& {(7)}& {(8)}& {(9)}& {(10)}& {(11)}& {(12)}& {(13)}&
{(14)}& {(15)}& {(16)}\\
\hline
 PSR   &$\dot{P}$& $P_1$& $P_2$& $P_3$& $\hat{P}_3$& $N$ & $r_p$ &
    $d$ & $v_d$ & $\frac{r_p}{h}$ & $\eta$ & $T_s$ & $T_i$ & $T_e$
    & $B_d$ & $B_s$ \\
    & $(10^{-15})$ & (s) & (deg) & $(P_1)$ & $(P_1)$ & & (m) & (m)
& (m/s) & & & ($10^6$K) &($10^6$~K) & ($10^6$K) &
    ($10^{12}$~G) & ($10^{12}$~G) \\
   \hline
B0943$+$10 & 3.25 & 1.098 & 10.5 & 1.86 & 37.35 & 20 & 95 & $\sim
80$ & $\sim 12$ & $\sim 7.4$ & 0.17 & 2.012 & 2.025 & 2.027 & 3.9
& $\sim 10$\\
B0809$+$74 & 0.17 & 1.29 & 11 & 11 & $>150$ & $\geq 15$ & 88 &
$\sim 73$ & $<2.4$ & $\sim 6$ & 0.032 & 0.955 & 0.956 & 0.957 &
0.94 & $\sim 7$\\
 B0826$-$34 & 1.0 & 1.85 & 26 & $\sim 1$ & $\sim 14$ & 14 & 73 & 33
& $\sim 8$ & $\sim 4.5$ & $\sim 0.36$ & 2.565 & 2.603 & 2.613
& 2.7 & $\sim 5$ \\
B2303$+$30 & 2.9 & 1.57 & 3 & $\sim 1.92$ & $\sim 23$ & 12 & 80 &
65 & $\sim 11$ & $\sim 5$ & $\sim 0.16$ & 1.854 & 1.864 & 1.865
& 3.3 & 9.6\\
B2319$+$60 & 7 & 2.26 & 16 & $\sim 7.8$ & $\sim 70$ & 9 & 66 & 52
& $\sim 2.1$ & $\sim 5$ & 0.071 & 1.236 & 1.239 & 1.239 & 7.9
& 13.6\\
B0031$-$07 & 0.4 & 0.94 & 20 & $\sim 6.8$ & $>34$ & $>5$ & 103 &
77 & $<15$ & $\sim 4$ & $>0.089$ & 1.551 & 1.556 & 1.557 & 1.2 &
7.2\\ \hline &&&&&&&&&&&&&&&& \\
&&&&&&&&&&&&&&&& \\ \multicolumn{9}{l}{(1)
$\dot{P}_{-15}=\dot{P}/10^{-15}$ --
period derivative} &&&&&&&& \\
\multicolumn{9}{l}{(2) $P_1$ -- pulsar period in seconds} &&&&&&&& \\
\multicolumn{9}{l}{(3) $P_2$ -- distance between driftbands in
degrees of longitude} &&&&&&&& \\
\multicolumn{6}{l}{(4) $P_3=1/f_3$ -- primary drifting period in
units of $P_1$} & \multicolumn{6}{l}{($f_3$ - frequency of drifting feature)}&&&& \\
\multicolumn{9}{l}{(5) $\hat{P}_3$ -- circulational (tertiary)
period in units of $P_1$ (Eq.~(A.13))} &&&&&&&& \\
\multicolumn{9}{l}{(6) $N=\hat{P}_3/P_3$ -- number of
circulating sparks (Eq.~(A.15))} &&&&&&&& \\
\multicolumn{9}{l}{(7) $r_p$ -- polar cap radius in meters
(Eq.~(A.4))} &&&&&&&& \\
\multicolumn{9}{l}{(8) $d$ -- circulation distance
($s=d/r_p\lesssim 1$)} &&&&&&&& \\
\multicolumn{9}{l}{(9) $v_d=2\pi d/\hat{P}_3$ -- circulation
speed in meters per second} &&&&&&&& \\
\multicolumn{9}{l}{(10) $r_p/h$ -- complexity parameter ($h$ - gap
height)}
&&&&&&&& \\
\multicolumn{9}{l}{(11) $\eta$ -- shielding factor (Eqs.~(5), (7)
and
(11))} &&&&&&&& \\
\multicolumn{9}{l}{(12) $T_s$ -- actual surface temperature in
$10^6$~K (Eq.~(10))} &&&&&&&& \\
\multicolumn{9}{l}{(13) $T_i$ -- ion critical temperature in
$10^6$~K (Eq.~(3) and
(4))} &&&&&&&& \\
\multicolumn{9}{l}{(14) $T_e$ -- electron critical temperature in
$10^6$~K
(Eq.~(5))} &&&&&&&& \\
\multicolumn{9}{l}{(15) $B_d$ -- dipolar surface magnetic field in
$10^{12}$~G (Eq.~(A.3))} &&&&&&&& \\
\multicolumn{9}{l}{(16) $B_s$ -- actual surface magnetic field in
$10^{12}$~G (Eq.~(A.2))} &&&&&&&& \\
\end{tabular}}

\end{document}